\documentclass[letterpaper,english,reprint, aps]{revtex4-1}
\usepackage[LGR,T1]{fontenc}
\usepackage[latin9]{inputenc}
\usepackage{color}
\usepackage{textcomp}
\usepackage{mathrsfs}
\usepackage{amsmath}
\usepackage{amssymb}
\usepackage{graphicx}

\makeatletter

\newcommand*\LyXThinSpace{\,\hspace{0pt}}
\pdfpageheight\paperheight
\pdfpagewidth\paperwidth

\DeclareRobustCommand{\greektext}{%
  \fontencoding{LGR}\selectfont\def\encodingdefault{LGR}}
\DeclareRobustCommand{\textgreek}[1]{\leavevmode{\greektext #1}}
\ProvideTextCommand{\~}{LGR}[1]{\char126#1}

\makeatother

\usepackage{babel}

\begin{document}

\title{Quantum coherence thermal transistors}

\author{Shanhe Su,$^{1}$ Yanchao Zhang,$^{1}$ Bjarne Andresen,$^{2}$ Jincan
Chen$^{1}$}
\email{jcchen@xmu.edu.cn}

\address{$^{1}$Department of Physics and Jiujiang Research Institute, Xiamen
University, Xiamen 361005, China.~\\
$^{2}$Niels Bohr Institute, University of Copenhagen, Universitetsparken
5, DK-2100 Copenhagen \O, Denmark.~\\
}

\date{\today}
\begin{abstract}
Coherent control of self-contained quantum systems offers the possibility
to fabricate smallest thermal transistors. The steady coherence created
by the delocalization of electronic excited states arouses nonlinear
heat transports in non-equilibrium environment. Applying this result
to a three-level quantum system, we show that quantum coherence gives
rise to negative differential thermal resistances, making the thermal
transistor suitable for thermal amplification. The results show that
quantum coherence facilitates efficient thermal signal processing
and can open a new field in the application of quantum thermal management
devices.

\begin{description}
\item [{PACS~numbers}] 05.90. +m, 05.70. \textendash a, 03.65.\textendash w,
51.30. +i
\end{description}
\end{abstract}
\maketitle

A thermal transistor, like its electronic counterpart, is capable of
implementing heat flux switching and modulating. The effects of negative
differential thermal resistance (NDTR) play a key role in the development
of thermal transistors \citep{key-1}. Classical dynamic descriptions utilizing
Frenkel-Kontorova lattices conclude that nonlinear lattices are the
origin of NDTR \citep{key-2,key-3}. Ben-Abdallah et al. introduced a distinct
type of thermal transistors based on the near-field radiative heat
transfer by evanescent thermal photons between bodies \citep{key-4}.
Joulain et al. first proposed a quantum thermal transistor with strong
coupling between the interacting spins, where the competition between
different decay channels makes the temperature dependence of the base
flux slow enough to obtain a high amplification \citep{key-5}. Zhang
et al. predicted that asymmetric Coulomb blockade in quantum-dot thermal
transistors would result in a NDTR \citep{key-6}. Stochastic fluctuations
in mesoscopic systems have been regarded as an alternate resource
for the fast switching of heat flows \citep{key-7}.

Recent studies showed that quantum coherence exhibits the ability to enhance
the efficiency of thermal converters, such as quantum heat engines
\citep{key-8,key-9,key-10} and artificial light-harvesting systems
\citep{key-11,key-12}. Interference between multiple transitions
in nonequilibrium environments enables us to generate non-vanishing
steady quantum coherence \citep{key-13,key-14}. Evidence is growing
that long-lived coherence boosts the transport of energy from light-harvesting
antennas to photosynthetic reaction centers \citep{key-15,key-16}.
The question arises whether quantum interference and coherence effects
could also induce nonlinear heat conduction and enhance the performance
of a thermal transistor.

Scovil and Schulz-DuBois originally proposed a three-level maser system
as an example of a Carnot engine and applied detailed balance ideas to obtain the
maser efficiency formula \citep{key-17}. Because the controlled (output)
thermal flux is normally higher than the controlling (input) thermal
flux, a thermal transistor is able to amplify or switch a small signal.
The amplification factor must be tailored to suit specific situations.
The Scovil and Schulz-DuBois maser model is not applicable for fabricating
thermal transistors, owing to the fact that its amplification factor
is simply a constant defined by the maser frequency relative to the pump frequency
\citep{key-18,key-19}. However, the coherent excitation-energy transfer
created by the delocalization of electronic excited states may aid
in the design of powerful thermal devices. Coherent control of a three-level
system (TLS) provides us a heuristic approach to better understand
the prime requirements for the occurrence of anomalous thermal conduction
in quantum systems.

In this paper we design a quantum thermal transistor consisting of a TLS coupled
to three separate baths. The dynamics of the system is derived by
considering the coupling between the two excited states. Steady-state
solutions will be used to prove that the coherent transitions between
the two excited states induce nonlinearity in nonequilibrium quantum
systems. Further analysis shows that quantum coherence gives rise
to a NDTR and helps improve the thermal amplification.

\begin{figure}
\includegraphics[scale=0.25]{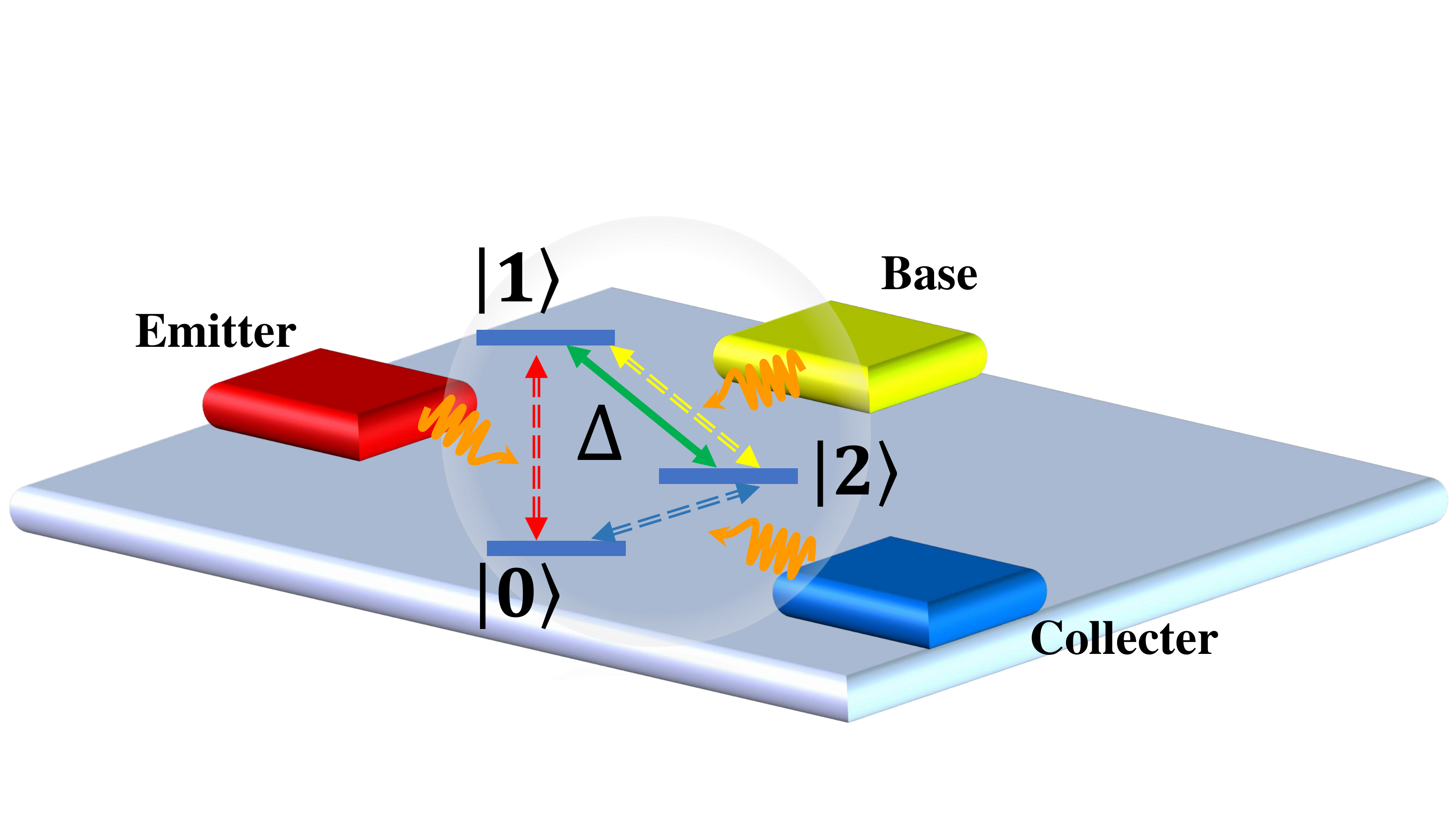}

\caption{{\footnotesize{}Schematic illustration of the quantum thermal
transistor composed of a three-level system (TLS) interacting with
three baths: its ground state $\left|0\right\rangle $ and excited
state $\left|1\right\rangle $ ($\left|2\right\rangle $) are coupled
with the emitter (collector); the excited states $\left|1\right\rangle $
and $\left|2\right\rangle $ are diagonal-coupled with the base; and
the coupling strength between $\left|1\right\rangle $ and $\left|2\right\rangle $
is characterized by $\Delta$.}}
\end{figure}

Figure 1 shows the TLS modeled by the Hamiltonian $H_{S}$ as
\begin{equation}
H_{S}=\sum_{i=0,1,2}\varepsilon_{i}\left|i\right\rangle \left\langle i\right|+\Delta(\left|1\right\rangle \left\langle 2\right|+\left|2\right\rangle \left\langle 1\right|),
\end{equation}
where $\varepsilon_{1}$ $\left(\varepsilon_{2}\right)$ gives the
energy level of the excited states in the molecules $\left|1\right\rangle $
$\left(\left|2\right\rangle \right)$, $\varepsilon_{0}$ denotes
the energy of the ground state $\left|0\right\rangle $ and is set
to zero, and $\Delta$ describes the excitonic coupling between states
$\left|1\right\rangle $ and $\left|2\right\rangle $. For the models
of biological light reactions, $\Delta$ occurs naturally as a consequence
of the intermolecular forces between two proximal optical dipoles
\citep{key-12,key-20}. In the presence of the dipole-dipole interaction,
the optically excited states become coherently delocalized. $\left|+\right\rangle =\cos\theta\left|1\right\rangle +\sin\theta\left|2\right\rangle $
and $\left|-\right\rangle =\sin\theta\left|1\right\rangle -\cos\theta\left|2\right\rangle $
are the usual eigenstates diagonalizing the subspace spanned by $\left|1\right\rangle $
and $\left|2\right\rangle $ with $\tan2\theta=2\Delta/\left(\varepsilon_{1}-\varepsilon_{2}\right)$.

The absorption of a photon from the emitter (E) causes an excitation
transfer from the ground state $\left|0\right\rangle $ to the state
$\left|1\right\rangle $, whereas phonons are emitted into the base
(B) by the transitions between $\left|1\right\rangle $ and $\left|2\right\rangle $.
The cycle is closed by the transition between $\left|2\right\rangle $
and $\left|0\right\rangle $, and the rest of the energy is released
as a photon to the collector (C). The Hamiltonians of the emitter, collector,
and base are $H_{i}=\sum_{k}\omega_{ik}a_{ik}^{\dagger}a_{ik}$ ($i=E$,
$C$, and $B$), where $a_{ik}^{\dagger}$ ($a_{ik}$) refers to the
creation (annihilation) operator of the bath mode $\omega_{ik}$.
The TLS couples to the emitter and the collector, each constituted of harmonic
oscillators, via coupling constants $g_{Ek}$ and $g_{Ck}$ in the
rotating wave approximation, where the corresponding Hamiltonians
are formally written as $H_{SE}=\sum_{k}\left(g_{Ek}^{\dagger}a_{Ek}\left|0\right\rangle \left\langle 1\right|+h.c.\right)$
and $H_{SC}=\sum_{k}\left(g_{Ck}^{\dagger}a_{Ck}\left|0\right\rangle \left\langle 2\right|+h.c.\right)$,
respectively. The output of the Scovil\textendash Schulz-DuBois maser
is a radiation field with a particular frequency, provided there is
population inversion between levels $\varepsilon_{1}$ and $\varepsilon_{2}$.
In this study, the two excited states are coupled with a thermal reservoir,
namely, the base. The interaction Hamiltonian of the system with the
base is described by

\begin{equation}
H_{SB}=(\left|1\right\rangle \left\langle 1\right|-\left|2\right\rangle \left\langle 2\right|)\sum_{k}g_{Bk}\left(a_{Bk}+a_{Bk}^{\dagger}\right).
\end{equation}
For a finite coupling $\Delta$, the base modeled by Eq. (2) induces
not only decoherence but also relaxation \citep{key-21}. The counterintuitive
effect of the energy exchange between the two excited states and the
dephasing bath becomes evident when the system operator coupled to
the base is replaced  by
\begin{align}
\left|1\right\rangle \left\langle 1\right| & =\cos\theta\cos\theta\left|+\right\rangle \left\langle +\right|+\sin\theta\sin\theta\left|-\right\rangle \left\langle -\right|\nonumber \\
 & +\sin\theta\cos\theta\left(\left|+\right\rangle \left\langle -\right|+\left|-\right\rangle \left\langle +\right|\right)
\end{align}
 and
\begin{align}
\left|2\right\rangle \left\langle 2\right| & =\sin\theta\sin\theta\left|+\right\rangle \left\langle +\right|+\cos\theta\cos\theta\left|-\right\rangle \left\langle -\right|\nonumber \\
 & -\cos\theta\sin\theta\left(\left|+\right\rangle \left\langle -\right|+\left|-\right\rangle \left\langle +\right|\right).
\end{align}
The first two operators in $\left|1\right\rangle \left\langle 1\right|$
and $\left|2\right\rangle \left\langle 2\right|$ describe the pure
dephasing of a two-level system, whereas the third term leads to the
energy exchange between the system and the base with an effective
coupling proportional to the product $\sin\theta\cos\theta$, i.e.,
\begin{equation}
H_{SB-eff}=2\sin\theta\cos\theta\left(\left|+\right\rangle \left\langle -\right|+\left|-\right\rangle \left\langle +\right|\right)\sum_{k}g_{k}\left(a_{k}+a_{k}^{\dagger}\right).
\end{equation}
In reality, the TLS can be realized in the photosynthesis process.
The pumping light, taking the sunlight photons for example, is considered
the high temperature emitter. The collector is formed by the surrounding
electromagnetic environment which models energy transfer to the
reaction center. The base provides the phonon modes coupled with the
excited states.

The TLS becomes irreversible due to the interaction with its surrounding
environment. Using the Born-Markov approximation, which involves
the assumptions that the environment is time independent and the environment
correlations decay rapidly in comparison to the typical time scale
of the system evolution \citep{key-22}, we get the quantum dynamics
of the system in $\hbar=1$ units, i.e.,

\begin{equation}
\frac{d\rho}{dt}=\text{\textminus}i[H_{S},\rho]+\mathcal{D_{\mathrm{\mathit{E}}}}[\rho]+\mathcal{D_{\mathit{B}}}[\rho]+\mathcal{D_{\mathit{C}}}[\rho].
\end{equation}
The operators $\mathcal{D_{\mathit{i}}}[\rho]$ ($i=E$, $B$, and
$C$) denote the dissipative Lindblad superoperators associated with
the emitter, base, and collector (Supplementary Eq. (S-1)), which take the
form

\begin{equation}
\mathcal{D_{\mathit{i}}}[\rho]=\sum_{v}\gamma_{i}\left(v\right)\left[A_{i}\left(v\right)\rho A_{i}^{\dagger}\left(v\right)-\frac{1}{2}\left\{ \rho,A_{i}^{\dagger}\left(v\right)A_{i}\left(v\right)\right\} \right],
\end{equation}
where $v=\varepsilon-\varepsilon^{\prime}$ is the energy difference
between two arbitrary eigenvalues of $H_{S}$, and $A_{i}\left(v\right)$
is the jump operator associated with the interaction between the system
and bath $i$. Considering a quantum bath consisting of harmonic oscillators,
we have the decay rate $\gamma_{i}\left(v\right)=\mathscr{\mathit{\Gamma}{}_{\mathit{i}}}\left(v\right)n_{i}\left(v\right)$
for $v<0$ and $\gamma_{i}\left(v\right)=\mathscr{\mathit{\Gamma}_{\mathit{i}}}(v)\left[1+n_{i}(v)\right]$
for $v>0$ , where $\mathscr{\mathit{\Gamma}_{\mathit{i}}}\left(v\right)$
labels the decoherence rate and is related to  the spectral density
of the bath, and $T_{i}$ is the temperature of bath $i$. The thermal
occupation number in a mode is written as $n_{i}(v)=1/\left[e^{v/\left(k_{B}T_{i}\right)}\text{\textminus}1\right]$.
The Boltzmann constant $k_{B}$ is set to unity in the following.

The steady-state populations and coherence of the open quantum system
are obtained by setting the left-hand side of Eq. (6) equal zero. Then the steady state energy
fluxes are determined by the average energy going through the TLS,
i.e.,
\begin{equation}
\overset{.}{E}(\text{\ensuremath{\infty}})=\underset{i=E,C,B}{\sum}\mathrm{Tr}\{H_{S}\mathcal{D_{\mathit{i}}}[\rho\left(\infty\right)]\}=\mathcal{\mathit{J}_{\mathit{E}}}+\mathit{J}_{\mathit{\mathit{C}}}\mathit{+}\mathit{J}_{\mathit{B}}=0
\end{equation}
which complies with the 1st law of thermodynamics. The heat fluxes
$\mathcal{\mathit{J}_{\mathit{E}}}$, $J_{\mathit{\mathrm{\mathit{C}}}}$,
and $\mathcal{\mathit{J}_{\mathit{B}}}$ are defined with respect
to their own dissipative operators. Thus,

\begin{align}
\mathcal{\mathit{J}_{\mathit{E}}} & =-\mathscr{\mathit{\Gamma}_{\mathit{E}}}\left(\varepsilon_{1}\right)\left(n_{E}+1\right)\left[\varepsilon_{1}\left(\rho_{1}-\frac{n_{E}}{n_{E}+1}\rho_{0}\right)+\varDelta\Re\left(\rho_{12}\right)\right]\nonumber \\
 & =J_{E1}+J_{E2},
\end{align}

\begin{align}
\mathcal{\mathit{J}_{\mathit{\mathrm{C}}}} & =-\mathscr{\mathit{\Gamma}{}_{\mathit{C}}}\left(\varepsilon_{2}\right)\left(n_{C}+1\right)\left[\varepsilon_{2}\left(\rho_{2}-\frac{n_{C}}{n_{C}+1}\rho_{0}\right)+\varDelta\Re\left(\rho_{12}\right)\right]\nonumber \\
 & =J_{C1}+J_{C2},
\end{align}
and

\begin{align}
\mathcal{\mathit{J}_{\mathit{B}}} & =-\mathscr{\mathit{\Gamma}_{\mathit{B}}}\left(\omega\right)\sin^{2}2\theta(2n_{B}+1)[\frac{\varepsilon_{1}-\varepsilon_{2}}{2}\left(\rho_{11}-\rho_{22}\right)\nonumber \\
 & +\frac{\sqrt{\left(\varepsilon_{1}-\varepsilon_{2}\right)^{2}/4+\varDelta^{2}}}{2n_{B}+1}+2\varDelta\Re\left(\rho_{12}\right)]=J_{B1}+J_{B2}.
\end{align}
The three heat fluxes are no longer linear functions of the rate
of the spontaneous emission, indicating that the symmetric property
is closely related to the base induced coherence of the excited states.
In Eqs. $(9) - (11)$, each heat flux is divided into two categories.
The terms $J_{i2}\left(i=E,C,B\right)$ are connected to the coherence
in the local basis, i.e., $\Re\left(\rho_{12}\right)$ (the real part
of $\rho_{12}$). $J_{i1}$ is the remainder components depending
on the populations of the TLS.

The thermodynamics of a TLS was originally proposed by Scovil and Schulz-DuBois
\citep{key-17}. Boukobza et al. obtained the Scovil\textendash Schulz-DuBois
maser efficiency formula when the TLS was operated as an amplifier
\citep{key-18,key-23,key-24}. The efficiency of the amplifier is
defined as the ratio of the output energy to the energy extracted
from the hot reservoir \citep{key-25}. In a nonequilibrium steady
state, the efficiency is a fixed value which equals $1-\left(\varepsilon_{2}-\varepsilon_{0}\right)/\left(\varepsilon_{1}-\varepsilon_{0}\right)$,
because all heat fluxes are linear functions of the same rate of
excitation. However, a thermal transistor is a thermal device used
to amplify or switch the thermal currents at the collector and the
emitter via a small change in the base heat flux or the base temperature.
Nonlinearity is the essential element needed to give rise to such thermal
amplification. For the purpose of flexible control of the thermal
currents, the characteristic functions of the TLS should not entirely
depend upon the energy level structure of the TLS.

A thermal amplifier requires a transistor with a high amplification
factor $\alpha_{E/C}$, which is defined as the instantaneous rate
of change of the emitter or collector heat flux to the heat flux
applied at the base. The quantum thermal transistor has fixed emitter
and collector temperatures $T_{E}$ and $T_{C}$ ($T_{E}>T_{C}$),
respectively. The fluxes $J_{E}$ and $J_{C}$ are controlled by $J_{B}$,
which can be adjusted by the base temperature $T_{B}$. Then the amplification
factor $\alpha_{E/C}$ explicitly reads

\begin{equation}
\alpha_{E/C}=\frac{\partial J_{E/C}}{\partial J_{B}}.
\end{equation}
Comparison of the slopes of the thermal currents is the key parameter
to find out whether the amplification effect exists. When $\left|\alpha_{E/C}\right|>1$,
a small change in $J_{B}$ stimulates a large variation in $J_{E}$
or $J_{C}$ and the thermal transistor effect appears. This implies
that a small change of the heat flux signal of the base would lead
to noticeable changes of the energy flowing through the emitter
and collector.

We consider heat fluxes from the baths into the TLS as positive.
As $T_{E}$ and $T_{C}$ are fixed values and $T_{B}$ is adjustable,
the thermal conductances of the three terminals are defined as

\begin{equation}
\sigma_{i}=-\frac{\partial J_{i}}{\partial T_{B}}=\sigma_{i1}+\sigma_{i2},
\end{equation}
where $\sigma_{ij}=-\frac{\partial J_{ij}}{\partial T_{B}}\left(i=E,C,B;j=1,2\right)$,
$\sigma_{i1}$ are the thermal conductances with respect
to the spontaneous emission, and $\sigma_{i2}$ are the thermal conductances
relying on the coherence $\Re\left(\rho_{12}\right)$. Using Eq.
(13), the amplification factor in Eq. (12) can be recast in terms of
$\sigma_{E}$ and $\sigma_{C}$, i.e.,
\begin{equation}
\alpha_{E/C}=-\frac{\sigma_{E/C}}{\sigma_{C}+\sigma_{E}}.
\end{equation}
The absolute value of the amplification factor $\left|\alpha_{E/C}\right|>1$
implies that one of the thermal conductances is negative, i.e., $\sigma_{C}<0$
or $\sigma_{E}<0$. This means that there exists a NDTR, and consequently,
the TLS can behave as a thermal transistor by controlling the heat
flow in analogy to the usual electric transistor.

In the following section, we need to explore the extent to which the
quantum nature of the TLS affects the thermal transistor. The formalism
obtained here will allow us to access how coherences can lead to a
NDTR and an enhancement of the amplification factor. To do so, the
thermal conductances and temperatures of the three baths are recast
in units of $\varDelta$. In the wide-band approximation, we write
the decoherence rates of the three terminals as $\mathscr{\mathit{\Gamma}_{\mathit{i}}}\left(v\right)=\mathscr{\mathit{\Gamma}_{\mathit{i}}}$
and the dephasing rate of the base as $\mathit{\gamma}_{B}\left(0\right)=\mathit{\gamma}_{0}$
.

Figure 2(a) shows the thermal conductances $\sigma_{i}$ of each terminal
as functions of the base temperature $T_{B}$. $\left|\sigma_{E}\right|$,
$\sigma_{C}$, and $\sigma_{B}$ decrease with $T_{B}$ at low temperature
and become constant as $T_{B}$ approaches $T_{E}$. As expected,
$\sigma_{B}$ remains lower than $\left|\sigma_{E}\right|$ and $\sigma_{C}$
over the whole range. A tiny change of the base heat flux $J_{B}$
or temperature $T_{B}$ is able to dramatically change the emitter
and collector thermal flows $J_{E}$ and $J_{C}$, leading to a noticeable
amplification effect. Similar to the decomposition of the thermal
fluxes, each thermal conductance can be divided into two separate
parts. Figures 2 (b) and (c) display the thermal conductances $\sigma_{i1}$
pertaining to the population distributions and to the coherence contributed
thermal conductances $\sigma_{i2}$ varying with the base temperature
$T_{B}$. $\sigma_{E1}$, $\sigma_{C1}$, and $\sigma_{B1}$ share
a magnitude close to each other, indicating that it is unlikely to
create an autonomous thermal amplifier without coherence. Quantum
coherence $\Re\left(\rho_{12}\right)$ exists {[}Fig. 2(d){]}, allowing
us to modify the thermodynamic behavior through the quantum control.
For the two thermal conductances $\sigma_{B1}$ and $\sigma_{B2}$
of the base, $\sigma_{B1}\text{>0}$ {[}Fig. 2 (b){]}, whereas $\sigma_{B2}$
originating from the coherence is negative {[}Fig. 2(c){]}, ensuring that
we achieve a vanishing $\sigma_{B}$ {[}Fig. 2(a){]}. Such a phenomenon
makes large thermal amplifications possible.

\begin{figure}
\includegraphics[scale=0.30]{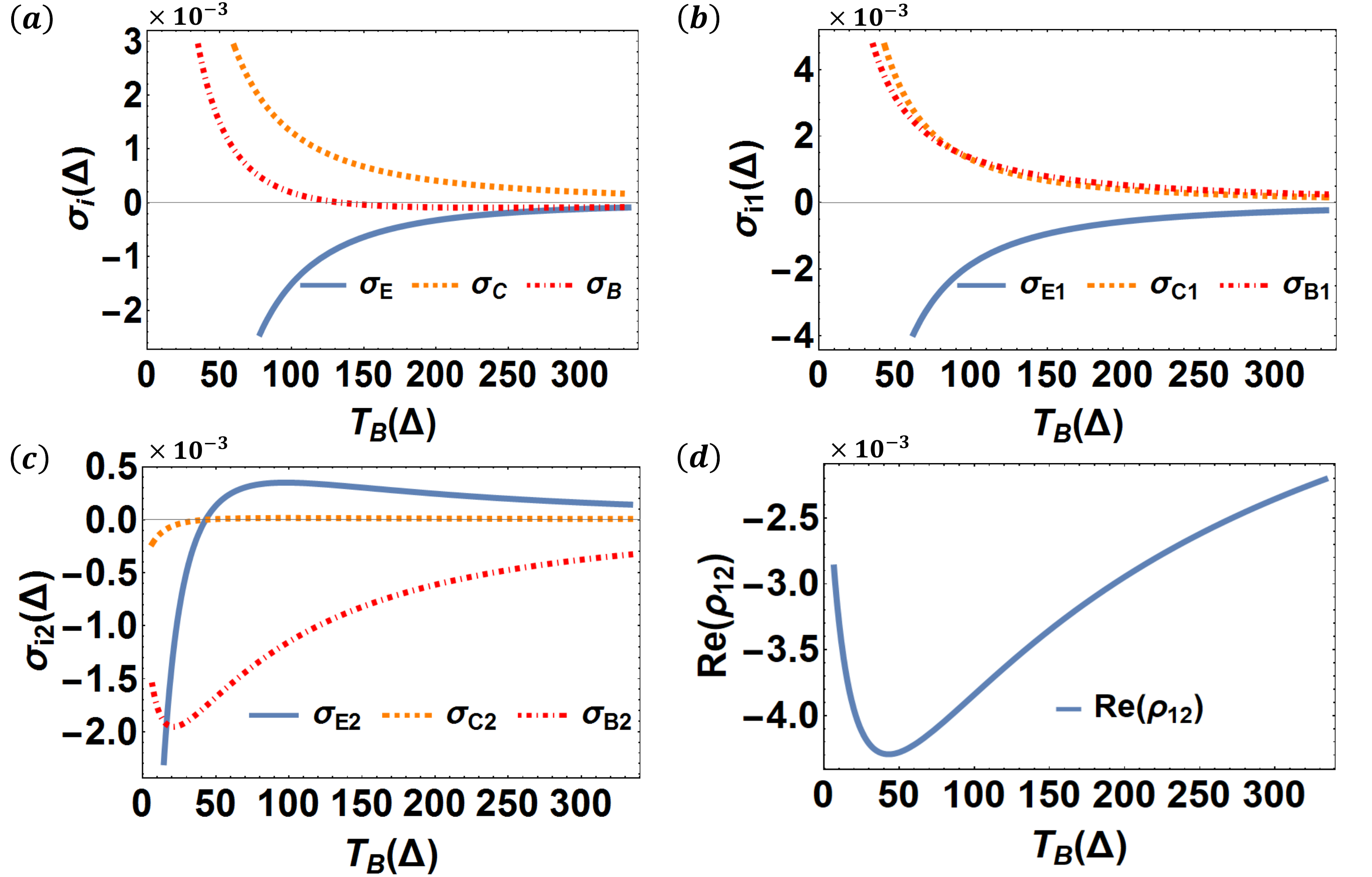}\caption{{\footnotesize{}(a) The overall thermal conductances $\sigma_{i}$;
(b) the thermal conductances $\sigma_{i1}$; (c) the thermal conductances
$\sigma_{i2}$; and (d) the real part of the coherence $\Re\left[\rho_{12}\right]$
versus the base temperature $T_{B}$. We choose the parameters in
units of $\varDelta$: $\mathscr{\mathit{\Gamma}_{\mathit{E}}}/\varDelta=\mathscr{\mathit{\Gamma}{}_{\mathit{C}}}/\varDelta=\mathscr{\mathit{\Gamma}_{\mathit{B}}}/\varDelta=\mathit{\gamma}_{0}/\varDelta\equiv1$,
$\varepsilon_{1}/\varDelta=10$, $\varepsilon_{2}/\varDelta=7$, $T_{E}=\varDelta/0.003$,
and $T_{C}=\varDelta/0.15$.}}
\end{figure}

The curves of the amplification factors $\alpha_{E}$ and $\alpha_{C}$
as functions of the base temperature $T_{B}$ are illustrated in Fig.
3. The amplification factors $\alpha_{E}$ and $\alpha_{C}$ are clearly
greater than 1 over a large range of $T_{B}$. As seen from Eq. (14),
these effects result from $\sigma_{E}<0$, which is similar to the
property of some electrical circuits and devices where an increase
in voltage across the overall assembly results in a decline in electric
current through it, i.e., negative differential conductance. Specifically, Fig. 3 shows that the amplification
factors diverge at $T_{B}=135.3\Delta$ due to the fact that the thermal
conductance of the base $\sigma_{B}=0$, induced by the quantum coherence.
Under these conditions, an infinitesimal change in $J_{B}$ makes
a considerable difference in $J_{E}$ and $J_{C}$.

\begin{figure}
\includegraphics[scale=0.25]{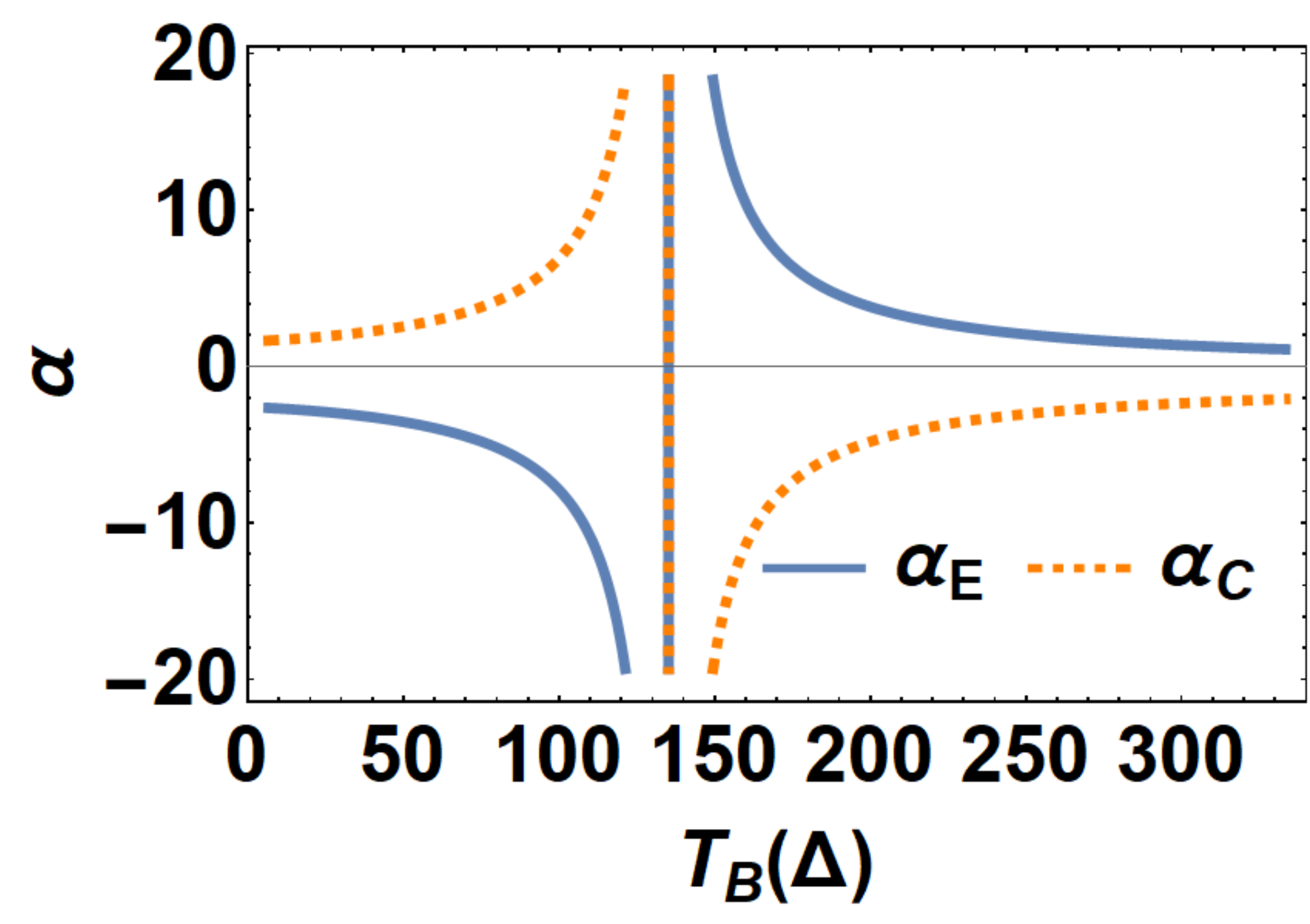}\caption{{\footnotesize{}The amplification factors $\alpha_{E}$ (solid line)
and $\alpha_{C}$ (dashed line) versus the base temperature $T_{B}$.
All parameters are the same as those used in Fig. 2}}
\end{figure}

Figures 4 and 5 reveal the influences of the decoherence rate $\mathscr{\mathit{\Gamma}_{\mathit{B}}}$
and the dephasing rate of the base $\mathit{\gamma}_{0}$ on the performance
of the thermal transistor. The base temperature $T_{B}=\varDelta/0.015$,
while the values of other parameters are the same as those used in
Fig. 2. The amplification factor $\alpha_{C}$ increases as $\mathscr{\mathit{\Gamma}_{\mathit{B}}}$
increases in the small-$\mathscr{\mathit{\Gamma}_{\mathit{B}}}$ regime
$(\mathscr{\mathit{\Gamma}_{\mathit{B}}}<1.287\text{\textgreek{D}})$,
but it decreases as $\mathscr{\mathit{\Gamma}_{\mathit{B}}}$ increases
in the large-$\mathscr{\mathit{\Gamma}_{\mathit{B}}}$ regime $(\mathscr{\mathit{\Gamma}_{\mathit{B}}}>1.287\text{\textgreek{D}})$,
while $\alpha_{C}$ tends to divergence for $\mathscr{\mathit{\Gamma}_{\mathit{B}}}\rightarrow1.287\text{\textgreek{D}}$.
The amplification factor $\alpha_{E}$ as a function of $\mathscr{\mathit{\Gamma}_{\mathit{B}}}$
has opposite signs. The decoherence rate $\mathscr{\mathit{\Gamma}_{\mathit{B}}}$
is an important parameter for building a desirable amplifier. As illustrated
in Figure 4(b), the thermal conductance $\sigma_{B}$ of the base
is the sum of $\sigma_{B1}$ and $\sigma_{B2}$. Once again, we observe
that $\sigma_{B1}$ is always positive, while the thermal conductance
relevant to the coherence effect $\sigma_{B2}\text{<0}$ leading to a cancellation of the sum when $\mathscr{\mathit{\Gamma}_{\mathit{B}}}\rightarrow1.287\text{\textgreek{D}}$. For the
same reason, the amplification factors diverge at $\mathscr{\mathit{\Gamma}_{\mathit{B}}}\rightarrow1.287\text{\textgreek{D}}$
when $\sigma_{B}=0$.

Coherence is maintained in a nonequilibrium steady state even
in the presence of the dephasing bath. However, a large dephasing
rate has a deleterious effect on the characteristics of the TLS thermal
transistor {[}Fig. 5(b){]}. Figure 5(a) shows that the absolute value
$\left|\rho_{12}\right|$ and the real part $\Re\left[\rho_{12}\right]$
of coherence are monotonically decreasing functions of $\mathit{\gamma}_{0}$, the decoherence rate of the base.
The pure-dephasing bath acting on the TLS induces the loss of steady
coherence, yielding smaller $\alpha_{E/C}$.

\begin{figure}
\includegraphics[scale=0.25]{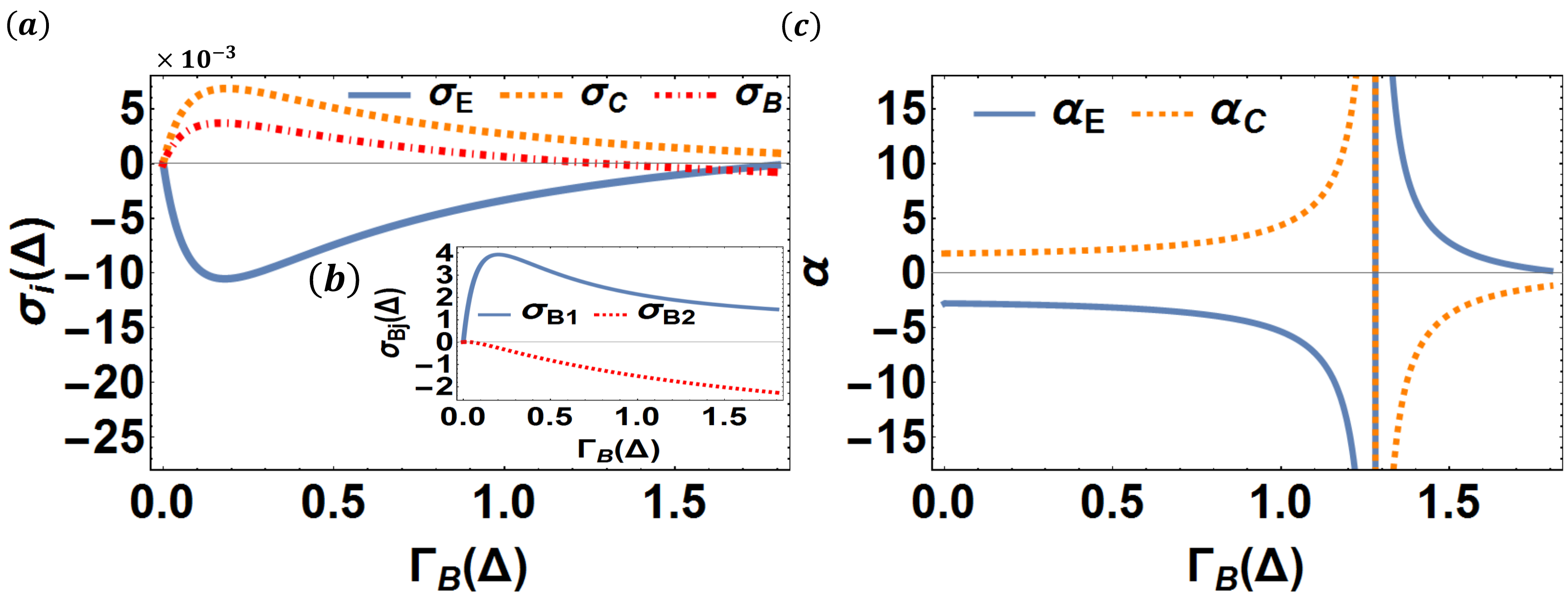}\caption{{\footnotesize{}(a) The overall thermal conductances $\sigma_{i}$;
(b) the thermal conductances of the base $\sigma_{Bj}$ (inset); and
(c) the amplification factors $\alpha_{E}$ (solid line) and $\alpha_{C}$
(dashed line) versus the decoherence rate of the base $\mathscr{\mathit{\Gamma}_{\mathit{B}}}$.}}
\end{figure}

\begin{figure}
\includegraphics[scale=0.25]{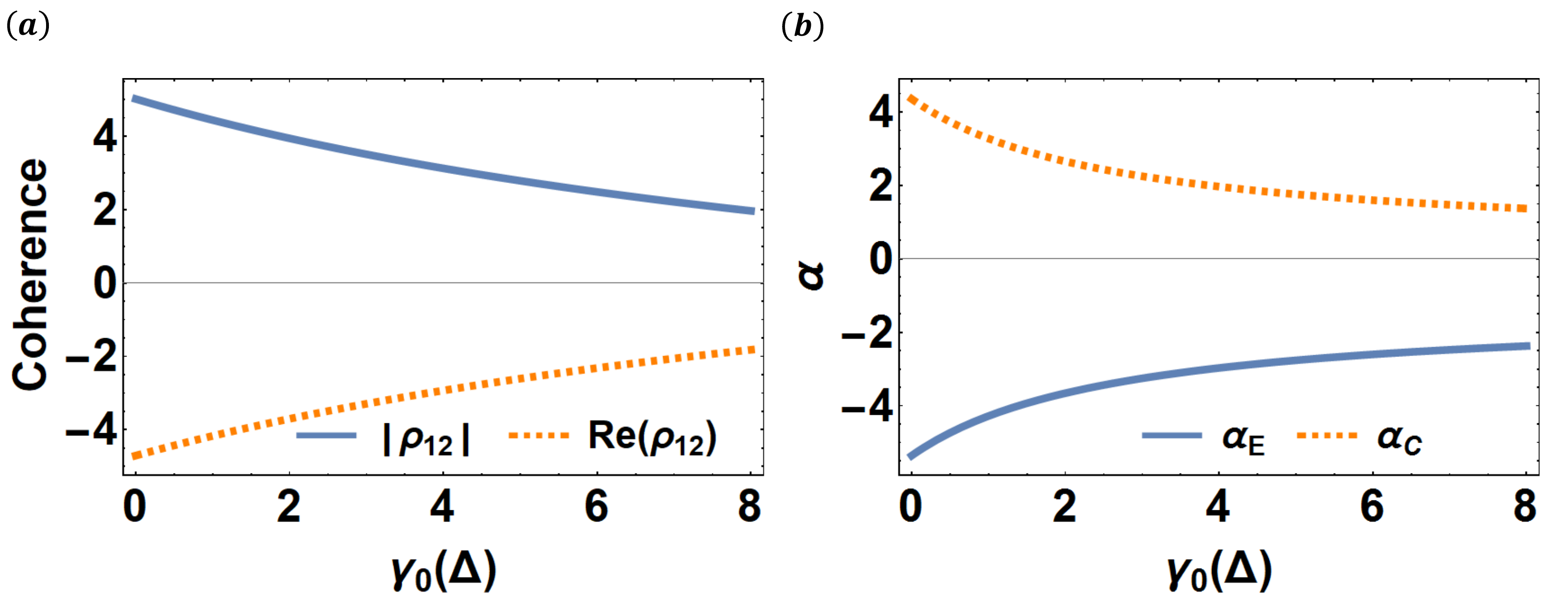}

\caption{{\footnotesize{}(a) The absolute value and the real part of coherence,
$\left|\rho_{12}\right|$ and $\Re\left[\rho_{12}\right]$, versus
the dephasing rate of the base $\mathit{\gamma}_{0}$. (b) The amplification
factors $\alpha_{E}$ (solid line) and $\alpha_{C}$ (dashed line)
versus the dephasing rate of the base $\mathit{\gamma}_{0}$.}}

\end{figure}

In summary, we build a TLS to analyze the effects of the dipole\textendash dipole
interaction and the dephasing on the energy transfer processes in
a thermal transistor. The coupling between the two excited states
of the TLS is capable of generating steady coherence in a nonequilibrium
environment, making the thermal fluxes behave nonlinearly. The coherence,
at the same time, gives rise to NTDR of the base. Quantum coherence
enables the thermal flow through the collector and emitter to be controlled
by a small change in the heat flux through the base. Such a thermal
transistor can amplify a small input signal as well as direct
heat to flow preferentially in one direction. The thermal transistor
effect can be significantly improved by optimizing the base temperature
and coherence rate or reducing the dephasing rate.

{\footnotesize{}We thank Dr. Dazhi Xu for helpful discussions. This
work has been supported by the National Natural Science Foundation
of China (Grant No. 11805159) and the Fundamental Research Fund for
the Central Universities (No. 20720180011).}{\footnotesize\par}

\end{document}